# Use of Digital Image Correlation to study the effect of temperature on the development of plastic instabilities in a semi-crystalline polymer

**Laurent Farge[1], Stéphane André[1], Julien Boisse[1]**

[1] Université de Lorraine, CNRS, LEMTA, F-54000 Nancy, France

Correspondence to: Laurent Farge (E-mail: *laurent.farge@univ-lorraine.fr*)

Postal adress:
Laboratoire d'Energétique et de Mécanique Théorique et Appliquée
LEMTA - UMR 7563
2, avenue de la forêt de Haye
BP 160
54504 Vandoeuvre-lès-Nancy
France

Abstract

*The plastic deformation processes that occur in a tensily deformed High Density Polyethylene specimen were studied from full-field strain and strain rate measurements obtained by 3D DIC (Digital Image Correlation). The tensile tests were performed every $10°C$ from room temperature to $120°C$. For temperatures below $60°C$, it is shown that the strain localization effect becomes less pronounced when the temperature increases. For temperatures higher than $60°C$, the material is found to exhibit double yielding behavior. By analyzing the DIC data in Lagrangian representation, it was possible to quantitatively highlight the strain localization effect that is specifically associated with the second yield. The second yield strain ($\varepsilon_{Y_2}$) was measured and appeared to be independent of temperature. For temperatures smaller than $60°C$, it was found that the threshold strain corresponding to $\varepsilon_{Y_2}$ also marks the onset of the deformation process phase during which the volume strain strongly increases. Based on previous studies of our research team and on literature we concluded that the critical strain $\varepsilon_{Y_2}$*

*corresponds to the onset of the lamellar morphology destruction. At high strain levels, the neck stabilization phase was shown to proceed according to a strain driven scheme characterized by threshold strains that are temperature independent. The experimental values of the threshold strain marking the onset of the stabilization phase are found to be in good agreement with those found using the Haward-Thackray model.*

Introduction

Several successive plastic regimes can be observed during tensile drawing of semi-crystalline polymers (SCPs) such as polyethylene [Gaucher-Miri (1996), Séguéla (2007)]. At temperatures comprised between the glass transition temperature ($T_g$) and the melting point ($T_m$), a first plastic instability called necking starts around the (first) yield point: strain localization occurs leading further to the progressive establishment of the neck. On a tensile curve, the necking onset is generally found by assuming that it occurs at the (first) force maximum. This can however be considered as an indirect detection since necking is essentially a feature of the deformation process and should be in principle detected through strain measurements. With further drawing, a "strain delocalization" effect, also called "neck stabilization", can be observed: In the neck region, the strain becomes progressively uniform and the strain rate tends to $0$. In parallel, the neck progressively propagates towards the nearly undeformed specimen regions. This is why the term "neck propagation" is also sometimes used to designate this stage of the plastic deformation process [Crist (2004)]. Besides this scenario, but only in some thermomechanical and/or microstructural configurations, an additional plastic regime called "second necking" can manifest itself during tensile drawing. When second necking occurs, it is also generally detected through the analysis of the force evolution: a second maximum corresponding to a second yield point appears on the force signal [Séguéla (1990), Brooks (1992), Séguéla (1994), Lucas (1995), Feijoo (1997)]. However, the second yield point is often partly blurred by the first one and then simply manifests itself through a shoulder on the force curve, especially in the case of High Density Polyethylenes (HDPEs) [Séguéla (1990), Brooks (1992), Séguéla (1994), Lucas (1995))]. For a given polyethylene grade, the appearance and development of a second yield is favored by increasing the temperature and by decreasing the strain rate [Séguéla (1990), Séguéla (1994), Lucas (1995))]. Furthermore, the polyethylenes with low crystallinity degrees are much more likely to exhibit double yielding behavior [Lucas (1995), Feijoo (1997)]. With in situ SAXS and WAXS (Small Angle X-ray Scattering and Wide Angle X-ray Scattering) experiments, it has been possible to differentiate the specific microstructure transformations that take place at the two yield points for polyethylene. At the first yield point, fine chain slip concomitantly with the martensitic transformation occurs while at the second yield point, coarse chain slip begins causing lamellar fragmentation [Sedighiamiri (2011), Schrauwen (2004), Butler (1997), Vickers (1995)].

At specimen scale, the strain localization effect is the dominant feature of plasticity during tensile deformation of SCPs. This leads several research teams to use full-field strain measurement techniques to study the inhomogeneous deformation that develops in these materials. Among these techniques, the 2D and 3D Digital Image Correlation (DIC) methods are by far the most used and have already demonstrated their potential to study the post-yield mechanical behavior of polymer materials for about 15 years [Parsons (2004), Parsons (2005), Fang (2006), Grytten(2009), Farge(2013)]. More specifically, based on DIC measurements performed at room temperature, it was possible to specifically highlight and analyze the plastic strain localization effect occurring in various polymer materials, for example: Polycarbonates nanocomposites [Christmann (2011)], Polypropylene/multiwall carbon nanotube nanocomposites [Ivanov(2014)], an epoxy resin [Poulain (2013)] or a HDPE [Ye (2015)]. In the last cited paper, the influence of the strain rate on the onset of the neck stabilization phase was also studied. However, not considering this exception, the DIC technique or any full-field strain measurement method was never used to study the detail of the successive phenomena involved in the full scenario of the plastic deformation process of SCPs. This requires to differentiate the successive strain localization effects that occur for a SCP exhibiting "double yielding" behavior and to study the steps leading to strain delocalization during neck stabilization.

Additionally, a very limited number of studies exists where DIC measurements were carried out on SCPs to study the influence of temperature on the plastic material behavior. In the case of a PA 12-based polymer subjected to tensile testing up to true strains smaller than $0.1$, the dependences of the material tensile properties on temperature ($-25°C$ to $50°C$) and on engineering strain rate ($0.00028\ s^{-1}$ to $9.4\ s^{-1}$) were characterized. The results were used to implement a temperature and strain rate dependent elasto-viscoplastic model [Serban (2013)]. Using Infra-Red thermography in addition to DIC, Johnsen et al. have also studied the influence of strain rate and temperature on the mechanical properties of a rubber-modified polypropylene and a cross-linked polyethylene for temperatures ranging from $-30°C$ to $25°C$ and for engineering strain rates ranging from $0.01\ s^{-1}$ to $1\ s^{-1}$ [Johnsen(2016,2017)]. It was notably found that the yield stress follows the Ree-Eyring flow theory [Johnsen(2017)]. As far as we know, in all the published works coupling DIC measurements to mechanical tests performed at various temperatures, only measurements obtained in the neck center, namely volume strains, true stress-true strain curves etc, were exploited. The specific influence of temperature on the strain localization/delocalization phenomena, an objective that basically requires the use of full-field strain measurements, was never specifically studied.

The objective of this paper is to study with 3D DIC measurements the influence of temperature on the plastic deformation process of a HDPE specimen subjected to tensile loading. The tests were performed at 11 levels of temperature ranging from $21°C$ to $120°C$. For $T > 60°C$, the material under study is shown to exhibit a "double yielding" behavior. In the major part of the study, we choose the Lagrangian representation, which allows for studying the strain localization/delocalization effects by evaluating the variations of the amount of substance included in the neck most deformed part. The temperature increment between two different tests was relatively limited: $10°C$. This made it possible to observe the progressive appearance and development of the second necking by measuring the strain localization effect which is specifically associated with it. It was found that the transition from necking (strain localization) to neck stabilization (strain delocalization) at the end of the test follows successive steps associated with characteristic strain levels that are temperature independent but specific to a well-defined state of the macromolecular network. Finally, we show that the values of these threshold strains can be satisfactorily predicted using the Haward-Thackray model.

# 1 Experimental section

## 1.1 Material

The SCP studied in this work is a HDPE (High Density PolyEthylene) provided by Röchling Engineering Plastics KG (grade ''500 Natural''). This material was manufactured by extrusion process and supplied in 6 mm thickness sheets. The supplier data sheet indicates molecular weight, density and melting temperature ($T_m$) of 500,000 g/mol, 0.95 g/cm3 and 135°C, respectively. A crystallinity index of about 68 wt% was measured by DSC (Differential Scanning Calorimetry). We used dog-bones shaped specimens with an initial cross-section $S_0$ of $6 \times 6\,mm^2$ in the specimen's central part. In figure 1, we show a picture of a specimen while it is being deformed. A detailed drawing of the undeformed specimen geometry is provided elsewhere (see figure 1 in Blaise et al. [Blaise (2012)]).

## 1.2 Tensile tests

The tensile tests were performed using the Bose® 3000 test machine equipped with a temperature chamber. The relative speed between the two grips was $0.02\,mms^{-1}$. The tensile test duration was about $1000\,s$. At the end of the test, the relative displacement of the grips was close to the maximum possible value for our tensile machine ($22\,mm$). The resulting true strains in the neck center ranged then from 1.6 to 1.85 for the different tests that were performed every $10^0 C$ from room temperature to $120^0$. The force $F$ was measured with a $3kN$ force sensor. The sampling rate for the recording of the force signal was $50\,ms$.

Strain fields measurements

The strain field measurements (see figure 1) were performed through a glass window placed on the front door of the temperature chamber using the 3D DIC equipment and the 2017 version of the Aramis® software provided by GOM company. The camera positions relatively to the specimen are the same as in Grytten et al. [Grytten(2009)]. This makes it possible to obtain strain measurements simultaneously on two perpendicular faces of the specimen: the main face, initially in the $(1,2)$ plane, and the lateral face in the $(1,3)$ plane. The $1$, $2$ and $3$ axis correspond to the specimen length, width and thickness directions respectively (see figure 1). The specimen deformation is characterized through the true (logarithmic) strain fields $\varepsilon_{11}$ (measured on the two specimen faces), $\varepsilon_{22}$ (measurement on the main face) and $\varepsilon_{33}$ (measurement on the lateral face). The strain fields were measured every one second. With the version of the Aramis software realeased in 2017, it was possible to obtain directly the strain fields measurements up to the end of the test. In our previous works, such high levels of investigated deformation required that we proceed in several stages to obtain proper correlations until the end of the tensile test [Ye (2015)].

The 3D DIC technique provides first displacement field measurements that are thereafter used to calculate strain fields. This allows for switching easily between the Eulerian (current configuration is the reference configuration) and Lagrangian (initial or undeformed configuration is the reference configuration) representations that are classically used in continuum mechanics. In figure 2, we show an example of the longitudinal strain $\varepsilon_{11}$ measured along path 1 (see figure 1) represented against both the Eulerian ($x_1$) and Lagrangian ($X_1$) coordinates. Due to the strain localization phenomenon (necking), the strain profiles exhibits relatively sharp peaks. The Maximum longitudinal strain at a given time is situated in the neck center and will be denoted $\varepsilon_{11}^M$ in the following. To characterize the size of the material domain where the strain localizes we will use the $\Delta X$ variable, which is the Full Width at Half Maximum (FWHM) of the longitudinal strain curve in Lagrangian representation (see figure 2). For the results presented in this paper, $\varepsilon_{11}(X_1)$ was measured along the tensile axis on the main face (see figure 2, path 1). $\Delta X$ is monitored as a function of two variables: the maximum strain $\varepsilon_{11}^M$ and the test temperature $T$. In the following, if $\Delta X$ is represented as a function of the test temperature $T$ and measured for every test at the same $\varepsilon_{11}^M$ strain, it will be denoted $\Delta X_{\varepsilon_{11}^M}(T)$. If it is measured during the same test, i.e at constant $T$, and represented as a function of $\varepsilon_{11}^M$, it is denoted $\Delta X_T\left(\varepsilon_{11}^M\right)$. By definition $\Delta X$ is a Lagrangian quantity corresponding to a size taken on the undeformed state and can be considered as a measure of the amount of substance that is included in the most deformed part of the neck, namely the part of the neck for which $\varepsilon_{11} > \varepsilon_{11}^M/2$. As a result, it is possible to highlight the strain localization/delocalization phenomena through the variations of the $\Delta X$ quantity.

Taking account of the great amount of data (11 separate tests associated to different temperatures × 1000 strain fields by test), it was necessary to develop automatic data processing methods to evaluate $\Delta X$ and the associated variable $\Delta X^c$ that will be defined later in section 2.2.

The strain rate was evaluated by calculating the material (Lagrangian) derivative from a centered difference formula:

$$\dot{\varepsilon}_{11}\left(X_1, t\right) = \frac{\varepsilon_{11}\left(X_1, t + \Delta t\right) - \varepsilon_{11}\left(X_1, t - \Delta t\right)}{2\Delta t} \qquad \text{eq. 1}$$

with $\Delta t = 10s$: one hundredth of the tensile test duration.

The volume strain was calculated in the central cross-section (neck center) and is given by:

$$\varepsilon_v = \varepsilon_{11}^M + \varepsilon_{22}^M + \varepsilon_{33}^M \qquad \text{eq. 2}$$

In eq. 2, $\varepsilon_{11}^M$, $\varepsilon_{22}^M$ and $\varepsilon_{33}^M$ were respectively evaluated by averaging $\varepsilon_{11}$, $\varepsilon_{22}$ and $\varepsilon_{33}$ along paths 2 (main face)+ path 3 (lateral face), path 2 and path3 (see figure 1). Strictly speaking, this equation is only valid if the deformation is homogeneous in the central cross-section where the measurement is made [Johnsen(2017)].

The true stress, ratio of the force on the current cross-section is given by:

$$\sigma_{11} = \frac{F}{S} = \frac{F}{S_0} \exp\left(-\varepsilon_{22}^M - \varepsilon_{33}^M\right) \qquad \text{eq. 3.}$$

$S$ is the current cross-section in the neck center.

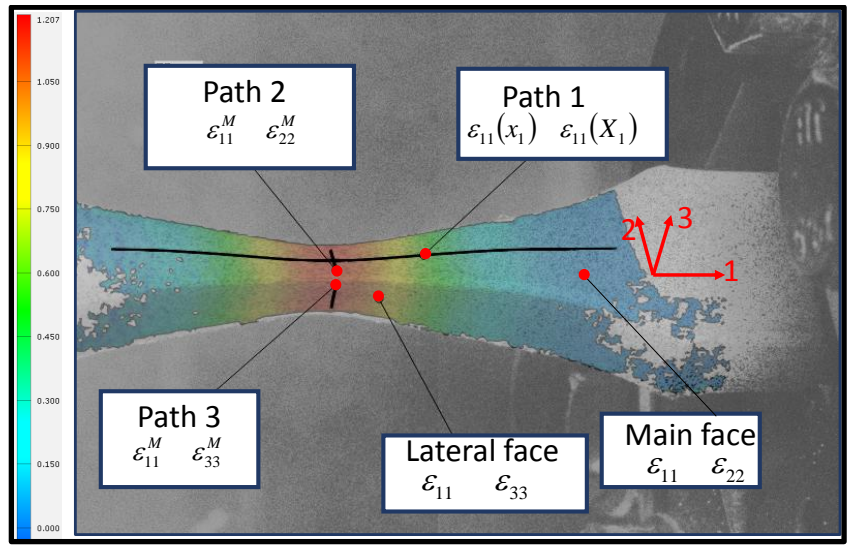

*Figure 1. Strain field measurements ($\varepsilon_{11}$) obtained by 3D DIC through the window of the temperature chamber for $T = 80°C$ and $\varepsilon_{11}^M = 1.21$.*

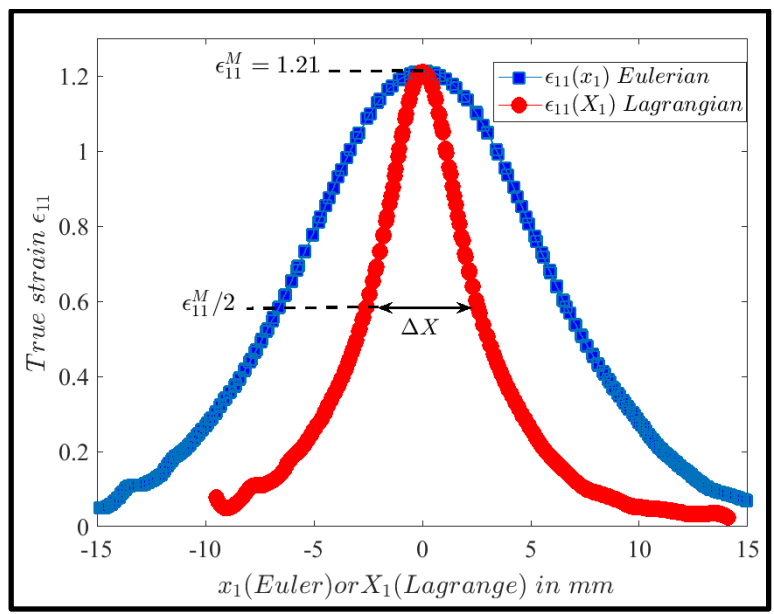

*Figure 2. Strain longitudinal profile along Path 1 of Figure 1 plotted against the Eulerian ($x_1$) and Lagrangian ($X_1$) coordinates.*

# 2 Results

## 2.1 Measurements in the neck center

In figure 3 we show the true stress-true strain $\left(\sigma_{11} - \varepsilon_{11}^M\right)$ curves obtained in the neck center for the different tests carried out at each temperature. The mechanical behavior of our polymeric material can be seen as greatly dependent on temperature.

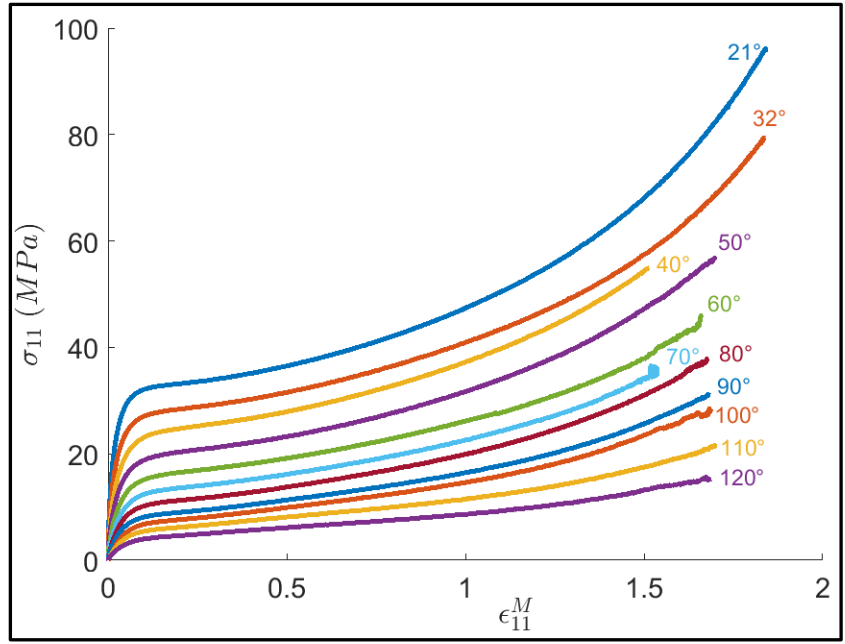

*Figure 3. Dependence of the True-Strain-True-Stress curves on the temperature.*

In figure 4, we represent in the same plot, the temporal evolutions of the normalized force $F/F^{\max}$ and of the maximum longitudinal strain $\varepsilon_{11}^M$. $F^{\max}$ is the force maximum, for example $F^{\max} = 1145\,N$ at $T = 21°C$ and $F^{\max} = 158\,N$ at $T = 21°C$. For the sake clarity, only a few curves are shown corresponding to $T = 21°C, 50°C, 60°C, 70°C, 80°C, 100°C$ and $120°C$ for $F/F^{\max}(t)$ as well as $T = 21°C$ and $120°C$ for $\varepsilon_{11}^M(t)$. The $F(t)$ and $\varepsilon_{11}^M(t)$ measured for all the test temperatures are provided in the supplementary file (figure S1). In table 1, we indicate the (first) yield stresses ($\sigma_{Y_1}$) and strains ($\varepsilon_{Y_1}$) simply obtained from the force maximum (Line 1 and 2) in that case. The drastic change of the polymer mechanical properties is illustrated by the variations of $\sigma_{Y_1}$ that nearly covers one order of magnitude.

We will focus first on the tests made at the lowest ($T = 21°C$) and highest temperatures ($T = 120°C$). At $T = 21°C$, the force curve has a "classical shape" with a maximum corresponding to the yield (denoted Y$_1$), which is approximately concomitant with the beginning of an increase phase for $\dot{\varepsilon}_{11}^M(t)$ (slope of the strain curve). Next, approximately for $\varepsilon_{11}^M \approx 0.7$, $\dot{\varepsilon}_{11}^M(t)$ begins to decrease before tending to 0 at the end of the test. On the other hand, for $T = 120°C$, the $\varepsilon_{11}^M(t)$ curve has a significantly different aspect: The first increase phase for $\dot{\varepsilon}_{11}^M(t)$ starting also approximately at the yield point (Y$_1$) is very short and finishes roughly at $\varepsilon_{11}^M \approx 0.25$. However, for about $\varepsilon_{11}^M \approx 0.5$, $\dot{\varepsilon}_{11}^M$ resumes to increase up to $\varepsilon_{11}^M \approx 1$ before tending to 0. As previously noticed and analyzed by Gaucher-Miri et al. [Gaucher-Miri (1996)], the $\varepsilon_{11}^M(t)$ curve observed at $T = 120°C$ with two distinct increase phases for $\dot{\varepsilon}_{11}^M$ evokes the successive development of two plastic regimes and therefore a "double yielding" behavior. This is confirmed by analyzing the force signal recorded at $T = 120°C$: after the first yield point, the $F/F^{max}$ curve clearly shows a shoulder that can be associated to the second yield [Séguéla (1990), Brooks (1992), Séguéla (1994), Lucas (1995))].

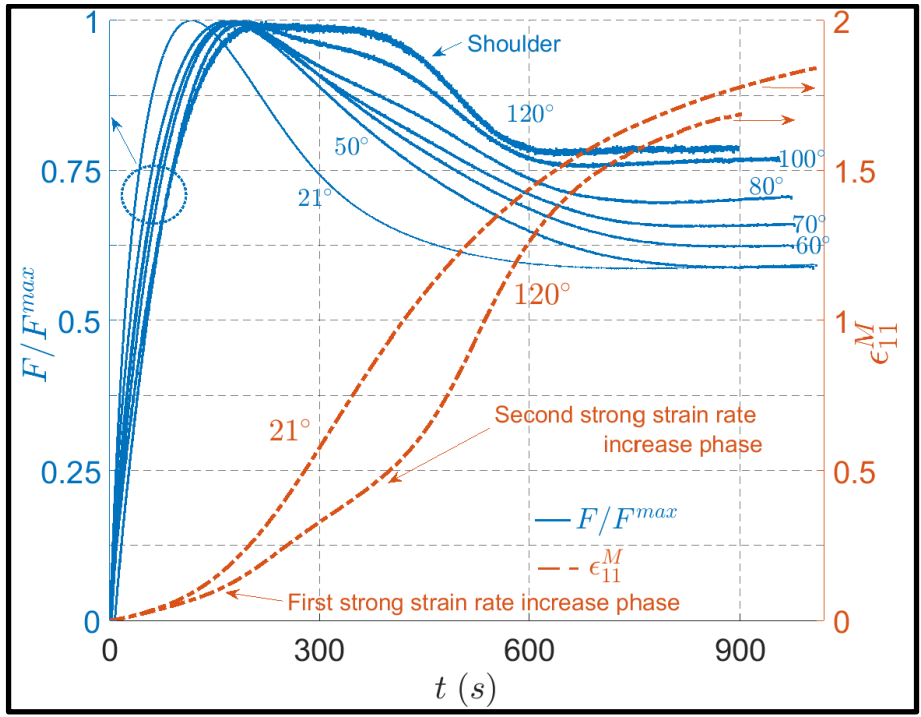

*Figure 4. Evolutions of $F/F^{max}$ (solide lines - left axis) and $\varepsilon_{11}^M$ (dotted lines - right axis) in the neck center for selected test temperatures.*

In order to evaluate an approximate transition temperature separating the "one yielding" and the "double yielding" behaviors, we also plotted the force and strain time derivatives, $\dot{F}$ and $\dot{\varepsilon}_{11}^{M}$, against $\varepsilon_{11}^{M}$ for all the temperatures (see the examples corresponding to $T = 21°C$, $50°C$, $60°C$, $70°C$, $80°C$ and $120°C$ in figure 5, and all the curves in figure S2). With simple mathematical reasoning, it is easy to check on the force curve that the shoulder associated to the second yield (Y$_2$) results in the appearance of a new maximum closely followed by a new minimum on the $\dot{F}$ signal. These new extrema are readily observable for the curves measured at $T \geq 80°C$. In the same temperature range but on the $\dot{\varepsilon}_{11}^{M}$ curve, the second yield appears also distinctly since it marks the onset of a second plastic regime during which the strain rate resumes to increase strongly. The characteristics of the "double yield" behavior keeps being discernable on the 70°C curves, in particular on the $\dot{\varepsilon}_{11}^{M}$ curve. The curve corresponding to $T = 60°C$ appears to correspond to a transition zone. The curves obtained for $T \leq 50°C$ do not exhibit the features characterizing the "double yield" behavior. The second yield strain values $\varepsilon_{Y_2}$ that are given in Table 1 (line 3) were found on the $\dot{\varepsilon}_{11}^{M}$ curves by determining the onset of the second increase (necking resumption). It is worth noting that the $\varepsilon_{Y_2}$ values are approximately the same for all the temperatures and are very close to $\varepsilon_{Y_2} \approx 0.4$. At $T = 70°C$, $\varepsilon_{Y_2}$ is $0.46$ and differs slightly from $\varepsilon_{Y_2} \approx 0.4$, but the measurement is difficult since the second necking effect is still not marked. On the force signal, the measurement of $\varepsilon_{Y_2}$ is impossible since the two force peaks associated to the two yield points overlap. The only way to determine precisely $\varepsilon_{Y_2}$ from the force evolution curves would be to deconvolute these two peaks as it is for instance currently done in the analysis of WAXS or Raman spectroscopy data by curve fitting. Unfortunately this could not be done for our force curves. For a tensile test, the force is not an intrinsic observable and there exists no shape function that are likely to describe reliably a peak force resulting from a yield point.

From the measurement of the maximum strain ($\varepsilon_{11}^{M}$) performed in the neck center, it was possible to quantitatively characterize the appearance of the first and possibly of the second yield, this latter being found to occur approximately for $\varepsilon_{11}^{M} = \varepsilon_{Y_2} \approx 0.4$. In the next section the strain localization effects specifically associated respectively with the first and with the second necking are studied through the analysis of the strain fields.

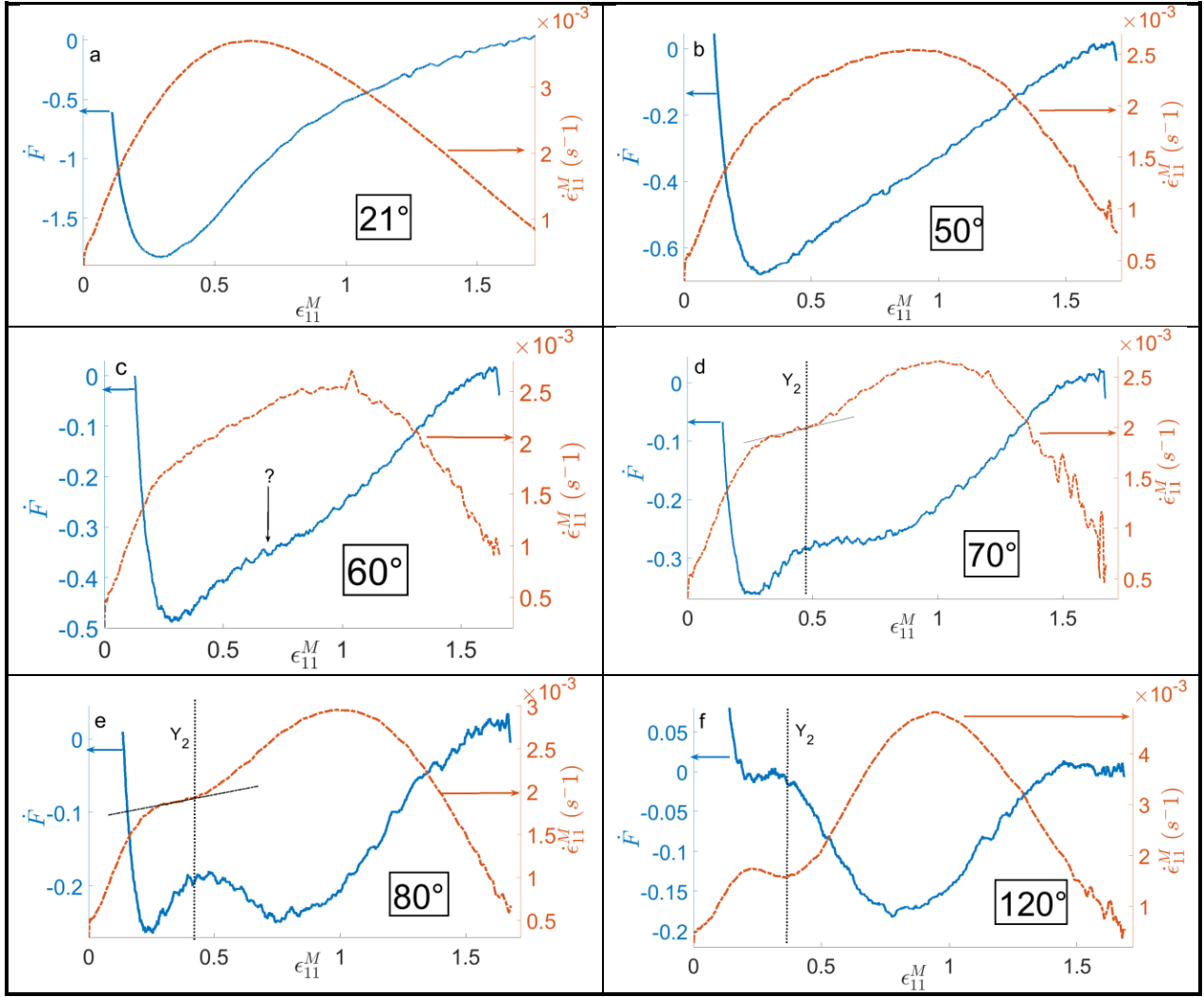

Figure 5. Evolution of $\dot{F}$ and $\dot{\varepsilon}_{11}^M$ against $\varepsilon_{11}^M$ a) $T = 21°C$, b) $T = 50°C$, c) $T = 60°C$, d) $T = 70°C$, e) $T = 80°C$ and f) $T = 120°C$.

| Temperature (C°) | 21 | 32 | 40 | 50 | 60 | 70 | 80 | 90 | 100 | 110 | 120 |
|---|---|---|---|---|---|---|---|---|---|---|---|
| 1st yield strain $\varepsilon_{Y_1}$ | 0.09 | 0.10 | 0.12 | 0.12 | 0.12 | 0.13 | 0.13 | 0.13 | 0.15 | 0.15 | 0.15 |
| 1st yield stress $\sigma_{Y_1}$ (MPa) | 31.8 | 27.0 | 23.3 | 19.3 | 15.6 | 12.9 | 10.7 | 8.7 | 7.3 | 5.8 | 4.4 |
| 2nd yield strain $\varepsilon_{Y_2}$ | × | × | × | × | × | 0.46 | 0.42 | 0.41 | 0.40 | 0.38 | 0.38 |

Table 1. First yield strains, first yield stresses and second yield strains obtained for the different test temperatures.

## 2.2 Strain field analysis: Study of the neck evolution in the Lagrangian representation

Temperature Influence on the strain localization effect associated with the first necking

In this part, we specifically study the influence of temperature on the development of the first necking. In figure 6, the $\Delta X_{0.4}$ variable is plotted against the test temperature. Defined in Lagrangian representation (see figure 2), $\Delta X_{0.4}$ can be seen as a measure of the amount of substance included in the most deformed part of the neck (part of the neck for which $\varepsilon_{11} > 0.4/2 = 0.2$) when the maximum central strain $\varepsilon_{11}^M = 0.4$ is the same for each of the specimen temperature. Based on the results of section 2.1, the value $\varepsilon_{11}^M = 0.4$ was chosen because it ensures that second necking has not started. In figure 6, the error bars were estimated by repeating the tests for certain temperatures. From $T = 21°C$ to $T = 60°C$, the $\Delta X_{0.4}$ variable increases. In this temperature range the strain localization effect becomes therefore less and less pronounced or, in other words, the necking becomes more and more diffuse when the temperature is increased (see the insert in figure 6). This was already observed [Séguéla (1994)] but never quantitatively measured. At higher temperatures, $T > 60°C$, $\Delta X_{0.4}$ was found to be nearly constant. This confirms that for $\varepsilon_{11}^M = 0.4$ the second necking, proved to be thermally activated [Séguéla (1990), Séguéla (1994), Lucas (1995)], has still not occurred.

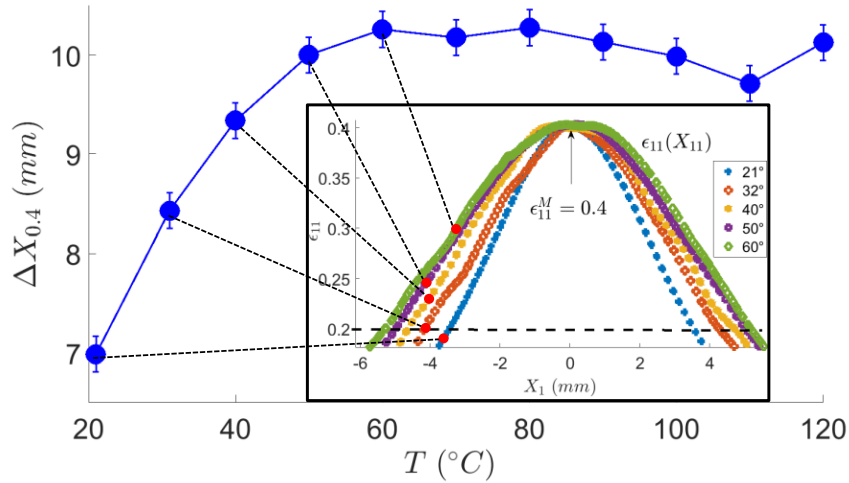

*Figure 6. Evolution of $\Delta X_{0.4}$ characterizing the material size of the neck for $\varepsilon_{11}^M = 0.4$ as a function of the temperature. The lower-right inset gives the $\varepsilon_{11}(X_1)$ profiles observed between $21°C$ and $60°C$ from which $\Delta X_{0.4}$ is measured.*

Temperature Influence on the strain localization effect associated with the second necking

To make evident the strain localization effect associated with the second necking, the same FWHM variable is calculated from the experimental data but on strain profile contrasts defined as: $\Delta\varepsilon_{11}(X_{11}) = B - A$, with:

- $B$: $\varepsilon_{11}(X_1)$ strain profile measured for $\varepsilon_{11}^M > 0.4$ (current state).
- $A$: $\varepsilon_{11}(X_1)$ strain profile measured for $\varepsilon_{11}^M = 0.4$ (reference state).

The reference state characterized by $\varepsilon_{11}^M = 0.4$ was chosen because it precedes the development of the second necking (see section 2.1). The FWHM measured on the $\Delta\varepsilon_{11}(X_{11})$ curve is denoted $\Delta X^c_{\varepsilon_{11}^M}$ with $\varepsilon_{11}^M$: strain maximum for the selected current state ($B$) profile. The superscript index $c$ was placed to remind that $\Delta X^c_{\varepsilon_{11}^M}$ is calculated from a strain profile **c**ontrast. Figure 7 illustrates the procedure for a test performed at $T = 90°C$ and for a current state corresponding to $\varepsilon_{11}^M = 1.1$. The FWHM corresponding to the contrast profile $\Delta\varepsilon_{11}(X_1)$ (figure 7) is $\Delta X^c_{1.1} = 3.54\,mm$, which is significantly smaller than that of the $B$ curve ($\Delta X_{1.1} = 4.97\,mm$). The shape of the reference $A$ curve ($\varepsilon_{11}^M = 0.4$), depends on temperature, which justifies this procedure to characterize the strain localization effect specifically associated with the appearance and development of the second necking. The interest of the Lagrangian representation is clear: the procedure makes sense only if the strain subtraction is calculated for the same material points. For all the curves shown in figure 7, the horizontal axis corresponds to the Lagrangian coordinate and the sizes $\Delta X^c_{1.1}$ and $\Delta X_{1.1}$ are both taken on the initial undeformed configuration. It should also be noted that the approach illustrated in figure 7 is based on the additivity property of the logarithmic strain.

In figure 8, we are now able to show the evolution of $\Delta X^c_{\varepsilon_{11}^M}$ as a function of temperature for $\varepsilon_{11}^M$ values ranging from $\varepsilon_{11}^M = 0.6$ to $\varepsilon_{11}^M = 1.3$. The development of the second necking with onset comprised between $T = 60°C$ and $T = 70°C$ is clearly evidenced by figure 8. In the $T \in [21\ 60]°C$ range, the $\Delta X^c_{\varepsilon_{11}^M}$ curves increase for all the $\varepsilon_{11}^M$ values: as previously observed (see figure 6), in this temperature range, the higher the temperature, the more diffuse the necking. On the other hand, in the $T \in [70\ 120]°C$ interval the $\Delta X^c_{\varepsilon_{11}^M}$ variable decreases which is due to the strain localization phenomenon resulting from the second necking appearing for $T \approx 60°C$ and intensifying as the test temperature increases.

In the $\varepsilon_{11}^M \in [0.6\ 1.3]$ range, all the curves are organized according to a logical order depending on the $\varepsilon_{11}^M$ value: the higher $\varepsilon_{11}^M$, the smaller $\Delta X_{\varepsilon_{11}^M}^c$. This shows that, in the considered strain range and for all the temperatures, the strain localization effect becomes more pronounced when $\varepsilon_{11}^M$ increases. The opposite trend is observed if $\varepsilon_{11}^M > 1.4$ : $\Delta X_{\varepsilon_{11}^M}^c$ increases when $\varepsilon_{11}^M$ increases (see figure S3). This curve sequencing inversion is related to the occurring of the neck stabilization phase as it will be seen in the following.

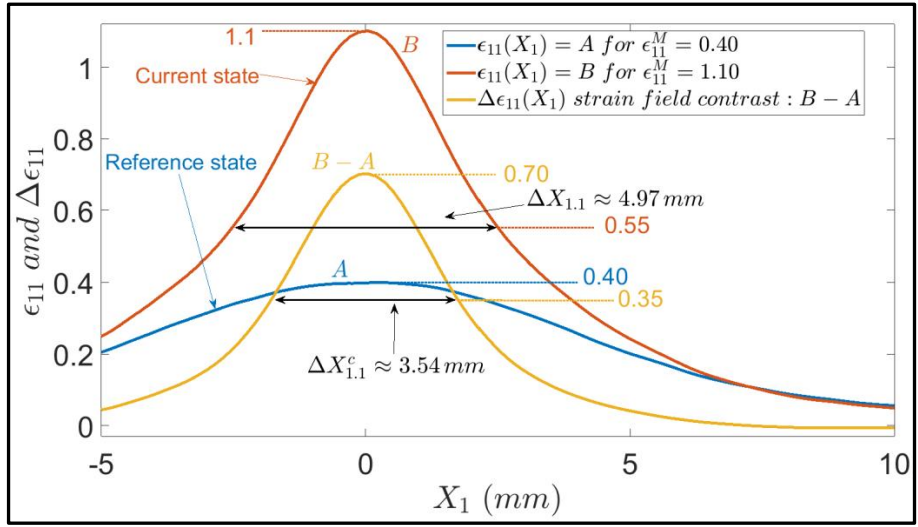

Figure 7. Illustration of the procedure used for obtaining $\Delta X_{\varepsilon_{11}^M}^c$ for $\varepsilon_{11}^M = 1.1$ and $T = 90°C$.

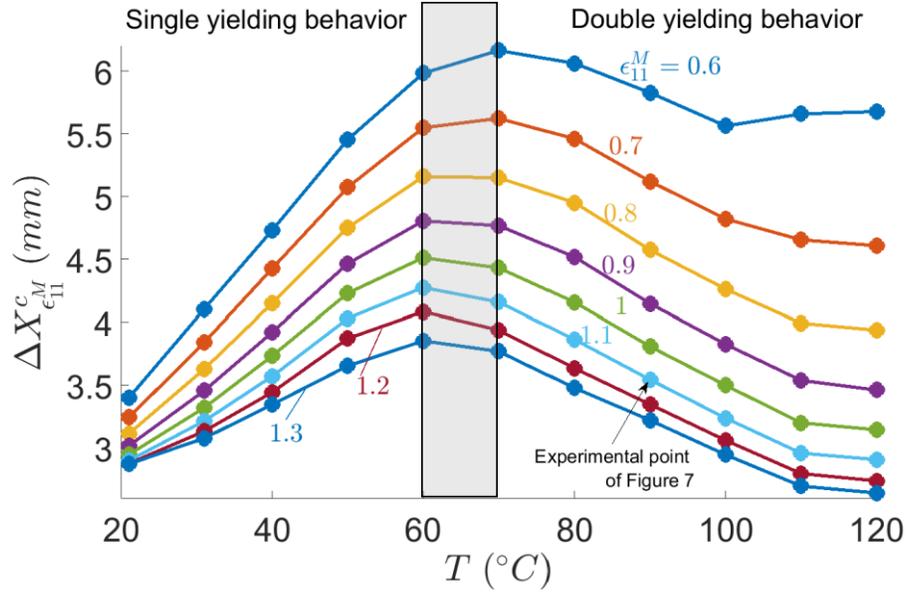

*Figure 8. Evolution of $\Delta X^c_{\varepsilon^M_{11}}$ as a function of temperature for $\varepsilon^M_{11} \in [0.6\ 1.3]$.*

Analysis of the $\Delta X_T \left( \varepsilon^M_{11} \right)$ curves

In this part we study the evolution of the $\Delta X_T$ variable as a function of $\varepsilon^M_{11}$. The plots corresponding to the two extreme temperatures, namely $T = 21°C$ and $T = 120°C$, are shown in figure 9 (left axis). Each of these curves are plotted with 1000 experimental points extracted from the corresponding recorded strain fields. At $T = 21°C$, it was previously shown that only one strain localization phase occurs during the deformation process; as a result, the $\Delta X_{21°C}$ curve has a relatively simple shape. On the other hand, the $\Delta X_{120°C}$ curve has a more complex form exhibiting two separate strain ranges during which the curve strongly decreases: the first can be observed for $\varepsilon^M_{11} < 0.3$ and the second is roughly centered around $\varepsilon^M_{11} = 0.7$. This causes the presence of a shoulder on the $\Delta X_{120°C}$ curve and it is evidently related to the successive development of two separate strain localization phases. On the same figure we also show the $\dfrac{d\left(\Delta X_{120°C}\right)}{d\varepsilon^M_{11}}$ plot corresponding to $T = 120°C$ (figure 9, right axis). For $\varepsilon^M_{11} < 0.4$, $\dfrac{d\left(\Delta X_{120°C}\right)}{d\varepsilon^M_{11}}$ is negative but increases: the strain localization effect beginning approximately at the first yield point is still in progress but becomes less and less pronounced. On the other hand, precisely at $\varepsilon_{11}^M \approx 0.4$, the decreasing phase for $\dfrac{d(\Delta X)}{d\varepsilon_{11}^M}$ begins. This can be attributed to the strain localization resumption (second yield point). Additionally, this confirms that the second yield point (Y$_2$) occurs approximately at $\varepsilon_{11}^M \approx 0.4$.

It can be seen in figure 9 that the $\Delta X_T$ curves measured for $T = 21°C$ and $T = 120°C$ reach a minimum value just before slightly increasing at the very end of the tests. Interestingly, the strains (denoted $\varepsilon_{NS}^1$ in the following) corresponding to the $\Delta X_T$ minimum are nearly the same at $T = 21°C$ and $T = 120°C$ ($\varepsilon_{NS}^1 \approx 1.5$). We wanted to check if this remains true for the tests carried out at the other temperatures. In the insert of figure 9, we show the curves corresponding to the evolution of the $\Delta X_T / (\Delta X_T)_{min}$ quantity for the 11 test temperatures of the study. $(\Delta X_T)_{min}$, minimum value of $\Delta X_T$, is a simple normalization factor, which makes the data easier to represent and to compare. The point is to notice that, for all the curves, the minimum value for $\Delta X_T$ is reached for about the same $\varepsilon_{NS}^1$ strain level. More precisely, the minimum is always comprised between $\varepsilon_{11}^M = 1.50$ and $\varepsilon_{11}^M = 1.63$. For all the test temperatures the $\varepsilon_{NS}^1$ strain values are given in Table 2. When $\varepsilon_{11}^M > \varepsilon_{NS}^1$, $\Delta X_T$ increases, which proves that the neck material size increases: the strain delocalization effect has begun. $\varepsilon_{NS}^1$ can therefore be seen as a direct indicator of the beginning of the neck stabilization phase. The lower script NS in $\varepsilon_{NS}^1$ stands for "Neck Stabilization". In the next section, the neck stabilization phase is studied in more detail through the analysis of the strain rate profiles measured along the longitudinal direction of the specimen.

| Temperature (°) | 21 | 32 | 40 | 50 | 60 | 70 | 80 | 90 | 100 | 110 | 120 |
|---|---|---|---|---|---|---|---|---|---|---|---|
| $\varepsilon_{NS}^1$ | 1.50 | 1.61 | 1.62 | 1.63 | 1.62 | 1.58 | 1.57 | 1.53 | 1.51 | 1.51 | 1.51 |
| $\varepsilon_{NS}^2$ | 1.40 | 1.48 | 1.51 | 1.52 | 1.50 | 1.47 | 1.47 | 1.44 | 1.42 | 1.42 | 1.41 |

*Table 2. threshold strains characterizing the beginning of the neck stabilization phase obtained through DIC measurements.*

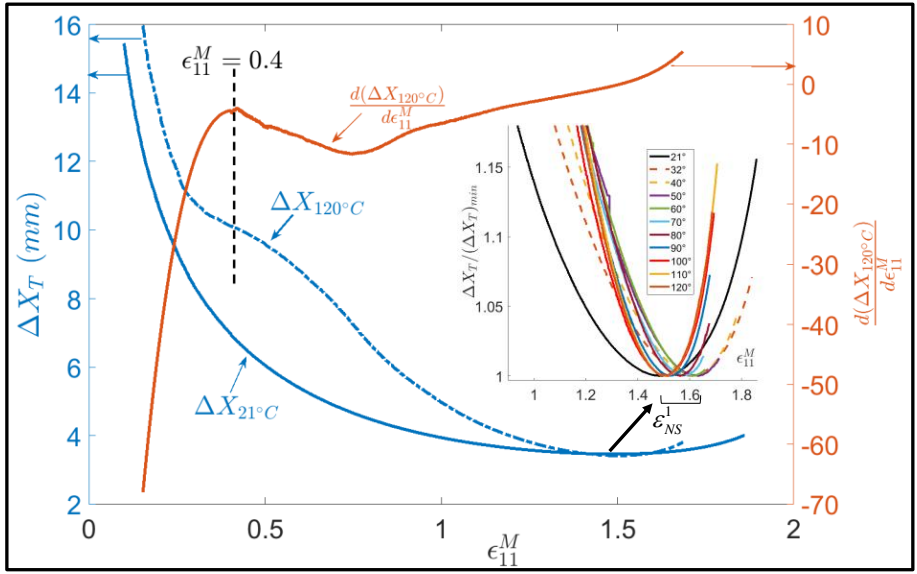

*Figure 9.* $\Delta X\left(\varepsilon_{11}^{M}\right)$ *curves for* $T = 21°C$ *and* $T = 120°C$. *The inset in the right part of the figure is an enlargement of the normalized quantity* $\Delta X_{T}/\left(\Delta X_{T}\right)_{min}$ *curves for all the test temperatures.*

Analysis of the strain rates $\dot{\varepsilon}_{11}(X_1)$ profiles

The longitudinal strain rate profiles $\dot{\varepsilon}_{11}(X_1)$ were calculated with eq. 1 for all the test temperatures. In figure 10, we show 7 examples of $\dot{\varepsilon}_{11}(X_1)$ profiles obtained during the test performed at $T = 90°C$ and corresponding to $t \in \{330, 500, 660, 720, 770, 830, 950\}\,s$. The strain profiles measured at the same time are shown in the supplementary file (figure S4). From the curves shown in figure 10, it is possible to differentiate two regimes: For $t \leq 720s$ ($\varepsilon_{11}^{M} = 1.426$), the strain rate profiles exhibit a unique maximum in the in the neck center ($t \in \{330, 500, 660, 720\}\,s$, R1 regime) while for higher times $t$ and strain values $\varepsilon_{11}^{M}$, two symmetrical maxima with respect to the specimen center progressively emerge (see figure 10 for $t \in \{770, 830, 950\}\,s$, R2 regime). The aim of figure 11 is to show the positions on the specimen of the two symmetrical peaks corresponding to the R2 regime when the deformation is in progress. In this figure, the $\varepsilon_{11}(x_1)$ and $\dot{\varepsilon}_{11}(x_1)$ curves measured for $T = 90°C$ and $\varepsilon_{11}^{M} = 1.67$ ($t = 950\,s$) are represented on the same plot and compared to a photography of the deformed specimen. In that case we choose the Eulerian coordinate $x_1$ for the $x$-axis, which allows a direct visual comparison between the curves and the specimen current state. However, it should be noted that, according to eq. 1, to obtain $\dot{\varepsilon}_{11}$ it is mandatory to calculate the $\varepsilon_{11}$ time derivative at constant $X_1$ (Lagrangian coordinate). It can be checked that the two symmetrical $\dot{\varepsilon}_{11}$ peaks are situated in the regions where $\varepsilon_{11}(x_1)$ strongly varies, i.e in the neck shoulders. Furthermore, in the Lagrangian representation of figure 10, it can be checked that the two peaks move apart from each other when the time increases: this straightforwardly shows that the shoulders progressively propagates towards material regions that were previously outside the neck. The transition from the R1 to the R2 regime was observed at all test temperatures and can be seen as a first step leading to neck stabilization. It is easy to conceive that this rapidly leads in an increase of the $\Delta X_T$ variable as it is observed at the end of tests shown in figure 9. In figure S5, we show that the threshold strain value $\varepsilon_{NS}^2$ that marks the transition from the R1 to the R2 regimes can be determined with a very good resolution: about 0.02. The $\varepsilon_{NS}^2$ values measured for all the test temperatures are indicated in table 2, line 3. It is worth noting that, for the large temperature range at which the tests were performed, $\varepsilon_{NS}^2$ remains approximately constant, comprised between $\varepsilon_{11}^M = 1.40$ and $\varepsilon_{11}^M = 1.52$, and slightly smaller than $\varepsilon_{NS}^1$.

Furthermore, when the test is nearly finished ($t = 950\ s$), the strain rate in the neck center becomes close to 0 and $\varepsilon_{11}^M$ tends to its final value $\varepsilon_{11}^{M, final} \approx 2$.

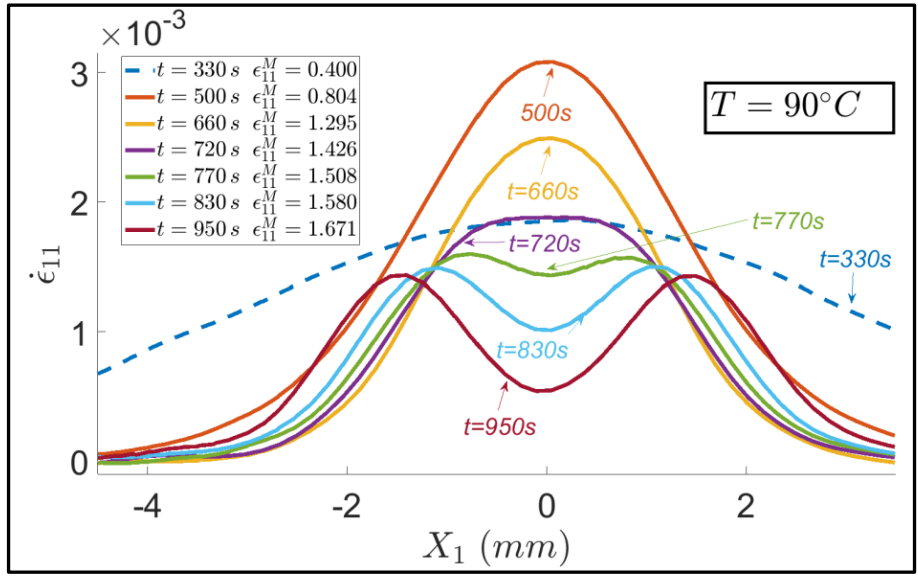

figure 10. Examples of $\dot{\varepsilon}_{11}(X_1)$ strain rate profiles for $T = 90°C$.

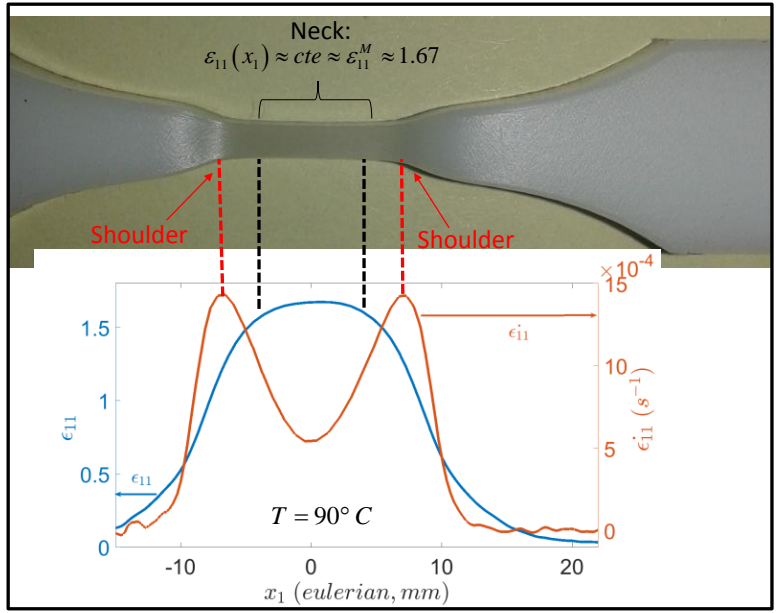

*Figure 11. $\varepsilon_{11}(x_1)$ and $\dot{\varepsilon}_{11}(x_1)$ profiles corresponding to $T = 90°C$ and $\varepsilon_{11}^M = 1.67$ ($t = 950\ s$). The use of the Eulerian representation allows for comparing the strain and strain rates profiles with the photography of a deformed specimen.*

## 3 Discussion

### 3.1 Neck Stabilization

Neck stabilization indicators

With our full-field strain measurements, it was possible to study the spatio-temporal evolution of the neck in Lagrangian representation. In particular, we highlighted two threshold strains (see table 2) that can be considered as direct indicators of the beginning of the neck stabilization stage:

1) $\varepsilon_{11}^M = \varepsilon_{NS}^2$, the strain rate maximum is no more situated in the neck center but in the neck shoulders.

2) $\varepsilon_{11}^M = \varepsilon_{NS}^1$, the amount of substance, characterized by the Lagrangian variable ($\Delta X_T$), included in the neck most deformed part begins to increase (see figure 9).

In the wide temperature range of our study ($T \in [21°C - 120°C]$), the material mechanical properties drastically change, but it is important to note that the $\varepsilon_{NS}^2$ and $\varepsilon_{NS}^1$ values remain nearly constant ( $\varepsilon_{NS}^2 \approx 1.4 - 1.5$ and $\varepsilon_{11}^M = \varepsilon_{NS}^1 \approx 1.5 - 1.6$, see table 2).

In the case of SCPs, it is generally considered that the limitation in extension of the molecular network is responsible for neck stabilization [Rietsch (1979)]. The nodes of this network consist initially of intercrystalline tie molecules and chain entanglements. At high strain levels, the lamellae are fragmented and the entangled amorphous network mainly governs the deformation process [Men(2003), Schrauwen (2004)]. It is assumed that the network's topological characteristics that are responsible for the limitation in material extensibility are mainly the chain lengths between two consecutive nodes [Séguéla (2007)]. The chain lengths between the nodes are obviously temperature independent and the same should be true for the strain level marking the beginning of neck stabilization. Therefore the experimental results we obtained here through the measurements of temperature quasi-independent $\varepsilon_{NS}^1$ and $\varepsilon_{NS}^2$ threshold strains, confirm that neck stabilization is a direct consequence of the limitation in extension of the molecular network.

Neck stabilization and Haward-Thackray model

The well-known Gaussian Haward-Thackray model (HT model) assumes that for relatively large strain levels the mechanical behavior of SCP's is dominated by the macromolecular network properties and can therefore be described with the laws of rubber elasticity [Haward (1968), Haward (1993)]. Bearing in mind that, as indicated in the previous section, neck stabilization is a network property, we wanted to test the ability of the HT model to predict the strain level at which this stage of the deformation process begins. In this model, the true stress is given by the so-called Gaussian equation:

$$\sigma_{11} = Y + G\left(\lambda^2 - 1/\lambda\right) \qquad \text{eq. 4}$$

$\lambda$ is the extension ratio that is directly related to the true strain: $\lambda = \exp(\varepsilon_{11})$. $Y$ is the extrapolated yield stress and $G$ is the network shear modulus. To derive the Gaussian equation, it is also necessary to assume that the material shows no dilatation during the deformation process, which will be shown to be not exactly true at $T = 21°C$ and $T = 40°C$.

Using the experimental data gathered in figure 3, it was found that the HT model works well for high strain levels. The values of $Y$ and $G$ were obtained by fitting the experimental curves $\sigma = f\left(\lambda^2 - 1/\lambda\right)$ with eq. 4 for $\lambda^2 - 1/\lambda > 12$ (see figure S6). The as-determined $Y$ and $G$ values are indicated in table 3 (line 2 and 3). At room temperature, it was checked that we obtain the same $Y$ and $G$ values for tests performed at constant strain rates.

Using the second Considère's criterion [Vincent(1960)] that assumes that a nominal stress minimum can be associated with the onset of neck stabilization, Haward has derived from eq. 4 a third order algebraic equation (eq. 7 in Haward's article [Haward (2007)]) giving $\lambda_{NS}^{HT}$. $\lambda_{NS}^{HT} = \exp(\varepsilon_{NS}^{HT})$ is the extension ratio marking the beginning of neck stabilization. Instead of the graphical solution method used by Harvard, a tractable analytical solution can be obtained and was used in the following:

$$\lambda_{NS}^{HT} = \exp(\varepsilon_{NS}^{HT}) = \left[\frac{(-2-\sqrt{\Delta})}{2}\right]^{1/3} + \left[\frac{(-2+\sqrt{\Delta})}{2}\right]^{1/3} \qquad \text{eq. 5}$$

With $\Delta = 4 + (4/27)c^3$ and $c = -Y/G$. Note that the two terms in eq. 5 are complex conjugate, which ensures that their sum is a real number. The $\varepsilon_{NS}^{HT}$ values for all test temperatures are indicated in the third line of table 3. In figure 12, we also show the temperature evolution of $G$, $\varepsilon_{NS}^{HT}$, $\varepsilon_{NS}^{1}$ and $\varepsilon_{NS}^{2}$. It can be checked that the $Y$ and $G$ values given by the HT model depend greatly on temperature. This is obviously related to the global alteration of the polymer mechanical properties occurring when the temperature increases. However, the first interesting result to point here is that the $\varepsilon_{NS}^{HT}$ threshold value obtained from the HT model remains roughly independent of temperature. The two first values are still slightly larger than the others ($T = 21°C$ and $T = 32°C$): the application of the Gaussian equation (eq 4) is maybe questionable in these two cases since it will be seen in the following that significant volume strain occurs during the deformation process at $T = 21°C$ and $T = 32°C$. Secondly, the agreement is pretty good between the $\varepsilon_{NS}^{HT}$ value and the $\varepsilon_{NS}^{1}$ and $\varepsilon_{NS}^{2}$ thresholds that were independently found through direct observation of the neck stabilization phase with full-field measurements (see figure 12). This proves that the HT model can be used satisfactorily by engineers to predict in a very simple way the onset of the neck stabilization phase.

| Temperature (°) | 21 | 32 | 40 | 50 | 60 | 70 | 80 | 90 | 100 | 110 | 120 |
|---|---|---|---|---|---|---|---|---|---|---|---|
| $G$ (MPa) | 1.48 | 1.17 | 1.25 | 1.07 | 0.89 | 0.82 | 0.80 | 0.66 | 0.62 | 0.41 | 0.30 |
| $Y$ (MPa) | 38.5 | 34.1 | 29.5 | 25.7 | 20.5 | 17.1 | 15.1 | 12.5 | 10.9 | 9.24 | 7.2 |
| $\varepsilon_{NS}^{HT}$ | 1.62 | 1.68 | 1.57 | 1.58 | 1.56 | 1.51 | 1.46 | 1.46 | 1.42 | 1.55 | 1.57 |

Table 3. $G$, $Y$ parameters of HT model and $\varepsilon_{NS}^{HT}$ threshold strain characterizing the beginning of the neck stabilization phase calculated with these parameters.

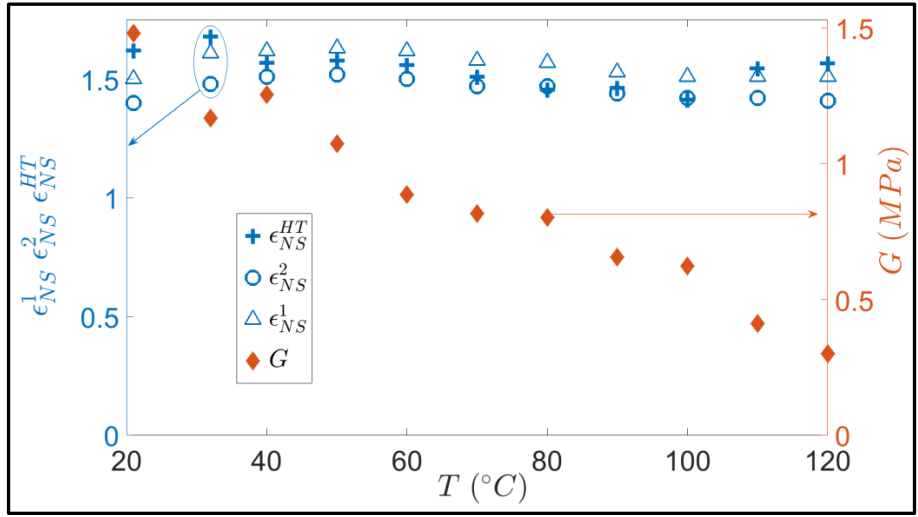

*Figure 12. Left axis: threshold strains $\varepsilon_{NS}^{1}$ (DIC measurement), $\varepsilon_{NS}^{2}$ (DIC measurement) and $\varepsilon_{NS}^{HT}$ (HT model) indicating the beginning the of neck stabilization phase. Right axis: network shear modulus $G$.*

Second necking and effect of lamellae fragmentation at the macro level

Previous SAXS-WAXS studies have shown that the second yield point can be associated to coarse chain slips leading to crystal fragmentation [Butler (1997), Vickers (1995)]. By analyzing our DIC measurements we found that the second necking is only observable if the tensile test is performed at $T \geq 60°C$. Above this temperature, we showed that it always begin at constant strain level $\varepsilon_{11}^{M} \approx 0.4$, which, as far as we know, was never observed before. Put together, these observations provide a novel evidence that, independently of temperature, the fragmentation of lamellae always occur at a constant critical strain, $\varepsilon_{11}^{M} \approx 0.4$ for HDPE [Hiss (1999), Hobeika (2000)]. However, the analysis is only valid for $T \geq 60°C$ and it would be interesting to see if it is possible to find a way to corroborate this result from our measurements for smaller temperatures. In that case the "double yield" behavior cannot be observed but, for HDPEs deformed at room temperature, it has already been shown that lamellae fragmentation is accompanied by a phase of strong irreversible volume strain [Addiego (2009), Ruihua (2008), Farge (2015)]. In figure 13, we show the volume strain evolutions ($\varepsilon_v$) measured during the tests performed at $T = 21°C$, $T = 32°C$ and $T = 40°C$, and calculated through eq. 2. The volume strain is found to decrease strongly when the temperature increases. This is due to the increase in mobility of the amorphous chains which makes impossible the nucleation and development of nanopores in the amorphous phase [Addiego (2006)]. Just after the yield ($\varepsilon_{11}^{M} \approx 0.1$), the volume strain becomes slightly negative for the tests performed at $T = 32°C$ and $T = 40°C$. Johnsen et al. have shown that this is a measurement bias [Johnsen (2017)]. Due to necking, the deformation is not homogeneous in the central cross section, which was confirmed by SAXS studies and Finite Element modelling [Farge (2018)]. As a result, eq. 2 is not strictly speaking applicable. In our case, this measurement bias makes the curves very difficult to interpret for $T \geq 40°C$ when the volume strain becomes very small. The three curves shown in figure 13 have overall the same trends and we will simply note that the phase of strong volume strain increase that can be associated to the lamellae fragmentation onset, begins precisely at the $\varepsilon_{11}^M \approx 0.4$ threshold strain. This is easy to check for $T = 21°C$ and $T = 32°C$ because the volume strain increase is well marked. This confirms that, independently of the temperature, lamellar fragmentation always occurs around $\varepsilon_{11}^M \approx 0.4$. In terms of microstructure deformation, this threshold strain can be interpreted as the maximum strain up to which the initial skeleton made of lamellae can deform while preserving its overall integrity.

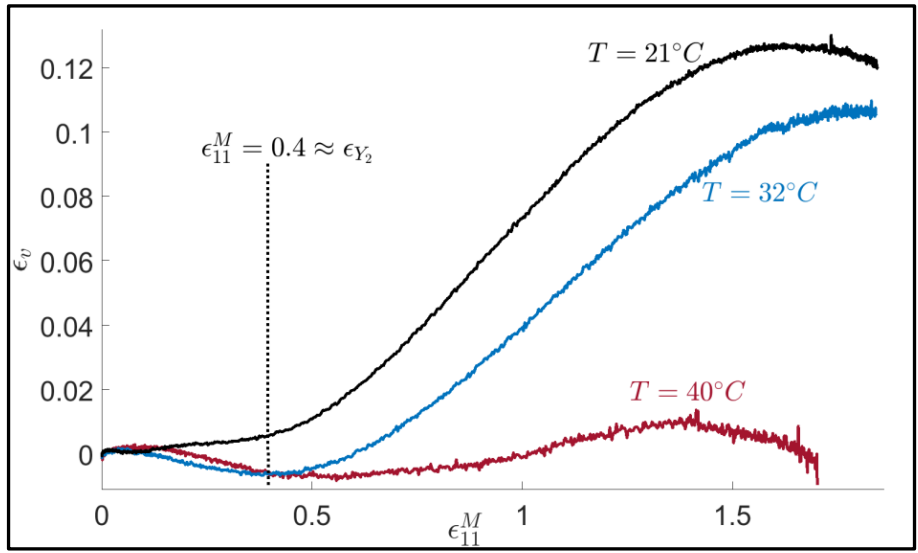

figure 13. Volume strain ($\varepsilon_v$) evolutions measured at $T = 21°C$, $T = 32°C$ and $T = 40°C$

Conclusion

The plastic deformation process was studied from strain and strain rate fields measured through 3D-DIC for a HDPE subjected to tensile tests performed at different temperatures ranging from $21°C$ to $120°C$. Of course it is expected that the material mechanical properties significantly change in this temperature range. But it was possible to measure and highlight the role of two critical strains that characterize the plastic deformation process of the polymer under study and that are independent of temperature. The first threshold strain $\varepsilon_{11}^{M} \approx 0.4$ corresponds to a second yield point for the tests performed at high temperatures. At low temperature, this threshold strain is associated to the onset of a strong volume strain increase phase. The second temperature-independent critical strain $\varepsilon_{11}^{M} \approx 1.4 - 1.6$ marks the onset of the neck stabilization phase. We also showed that this strain threshold can be found independently by applying the Haward-Thackray model.

The existence of these strain thresholds can be simply explained by assuming that they are related to limits in extensibility of the networks that, independently of temperature, structure the semi-crystalline polymer under study. The first threshold strain can be interpreted as corresponding to the maximum deformation to which can be subjected the lamellar network while keeping its integrity, i.e without fragmentation of the lamellae. The second threshold strain can be associated to the extensibility limit of the macromolecular network.

Schrauwen, B. A., Janssen, R. P., Govaert, L. E., & Meijer, H. E. (2004). Intrinsic deformation behavior of semicrystalline polymers. *Macromolecules*, *37*(16), 6069-6078.

Sedighiamiri, A., Govaert, L. E., & Van Dommelen, J. A. W. (2011). Micromechanical modeling of the deformation kinetics of semicrystalline polymers. *Journal of Polymer Science Part B: Polymer Physics*, *49*(18), 1297-1310.

Seguela, R., & Rietsch, F. (1990). Double yield point in polyethylene under tensile loading. *Journal of materials science letters*, *9*(1), 46-47.

Seguela, R., & Darras, O. (1994). Phenomenological aspects of the double yield of polyethylene and related copolymers under tensile loading. *Journal of materials science*, *29*(20), 5342-5352.

Séguéla, R. (2007). On the Natural Draw Ratio of Semi-Crystalline Polymers: Review of the Mechanical, Physical and Molecular Aspects. *Macromolecular Materials and Engineering*, *292*(3), 235-244.

Şerban, D. A., Weber, G., Marşavina, L., Silberschmidt, V. V., & Hufenbach, W. (2013). Tensile properties of semi-crystalline thermoplastic polymers: Effects of temperature and strain rates. *Polymer Testing*, *32*(2), 413-425.

Vickers, M. E., & Fischer, H. (1995). Real-time in situ X-ray diffraction study of polyethylene deformation. *Polymer*, *36*(13), 2667-2670.

Vincent, P. I. (1960). The necking and cold-drawing of rigid plastics. *Polymer*, *1*, 7-19.

Ye, J., André, S., & Farge, L. (2015). Kinematic study of necking in a semi-crystalline polymer through 3D Digital Image Correlation. *International Journal of Solids and Structures*, *59*, 58-72.
28

# Supplementary file

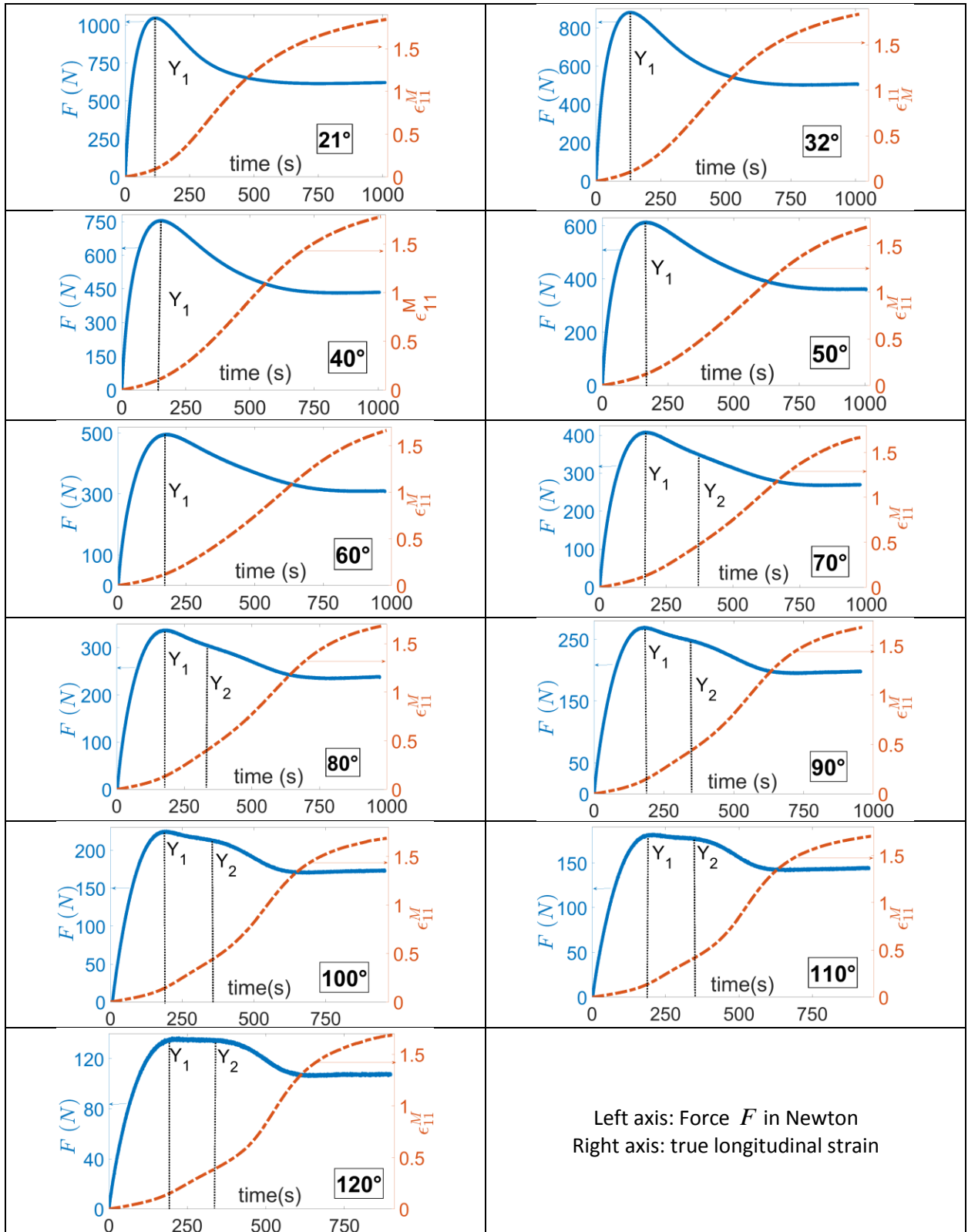

*Figure S1: Evolutions of $F(t)$ and $\varepsilon_{11}^M(t)$ for all test temperatures.*

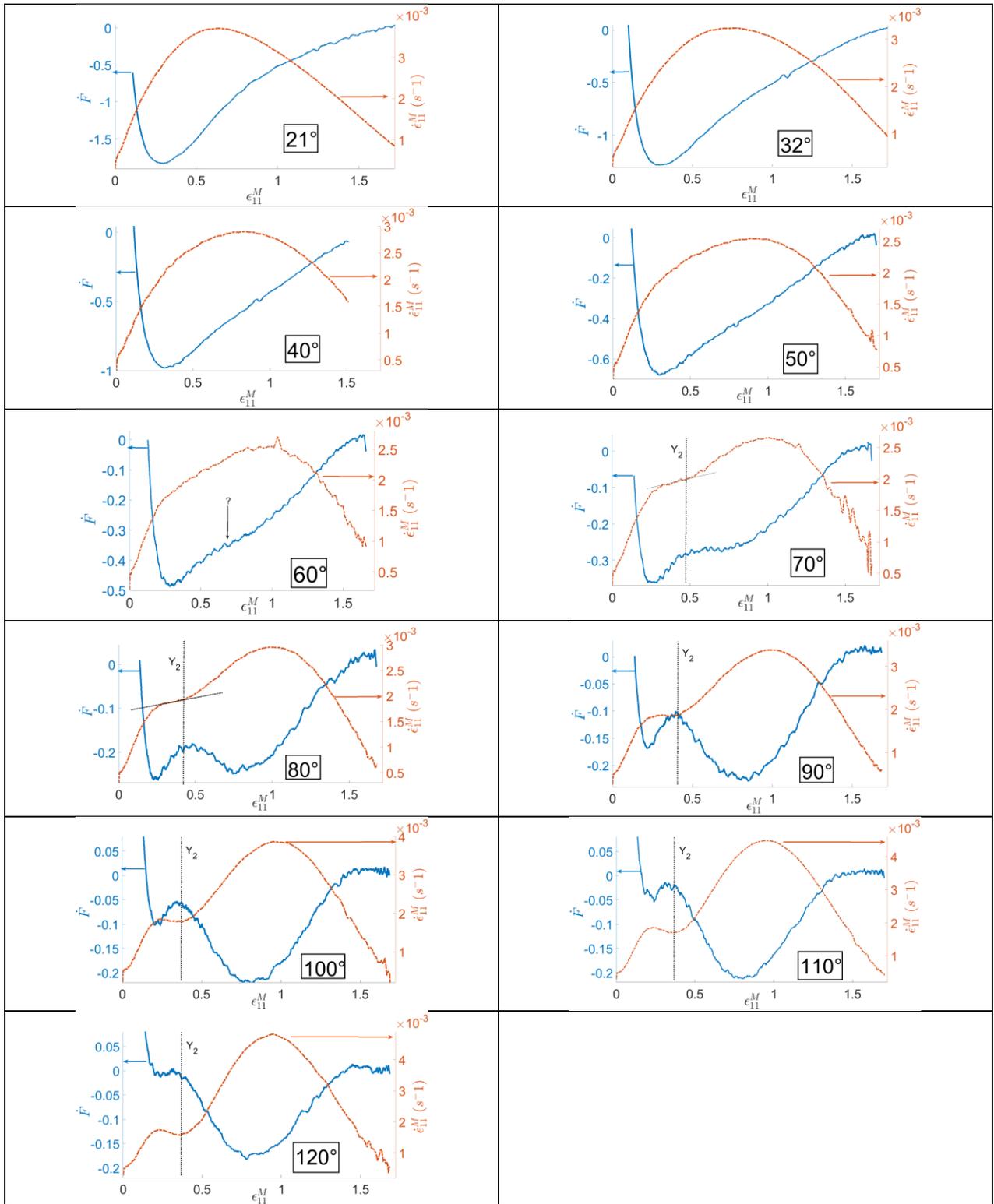

*Figure S2.* Evolution of $\dot{F}$ and $\dot{\varepsilon}_{11}^{M}$ against $\varepsilon_{11}^{M}$ for all the test temperatures.

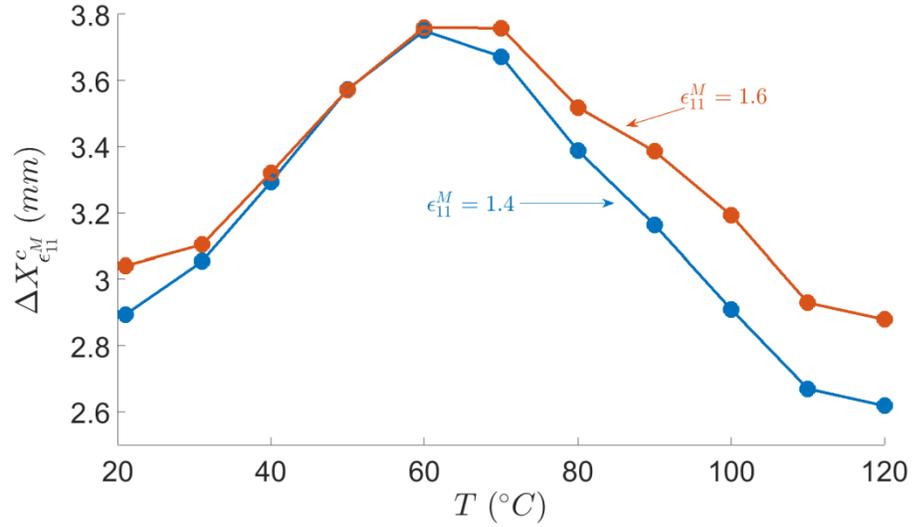

*Figure S3. Evolution of $\Delta X^c_{\varepsilon^M_{11}}$ for $\varepsilon^M_{11} = 1.4$ and $\varepsilon^M_{11} = 1.6$: compared to figure7, the curve order has changed. The curve obtained at the larger strain level $\varepsilon^M_{11} = 1.6$ is above that measured at $\varepsilon^M_{11} = 1.4$*

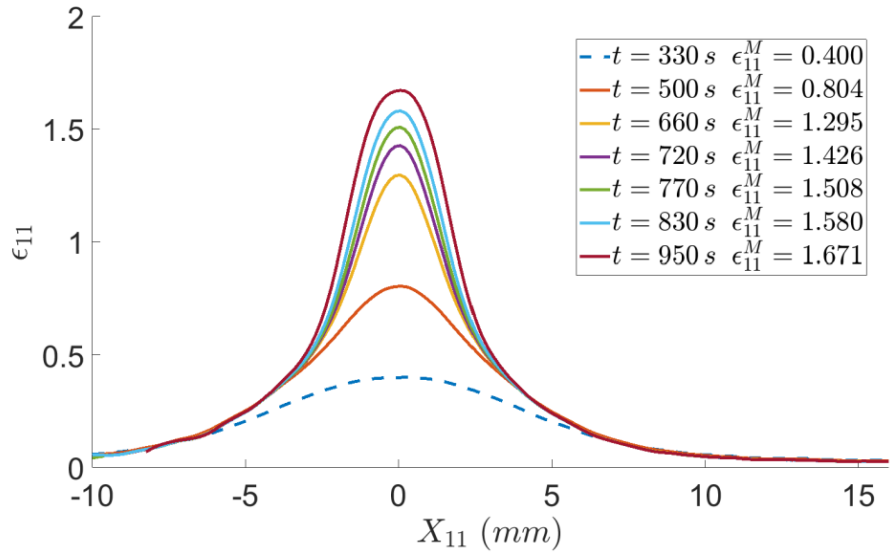

*Figure S4. Strain profiles measured at the same time as the strain rates shown in figure 9.*

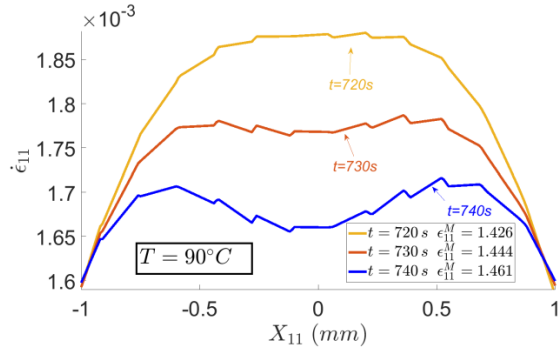

*Figure S5 Three $\dot{\varepsilon}_{11}(X_1)$ strain rate profiles separated by short time intervals taken around the transition between R1 and R2 regimes. This shows that the strain threshold marking the transitions from the R1 regime to the R2 regime can be accurately determined.*

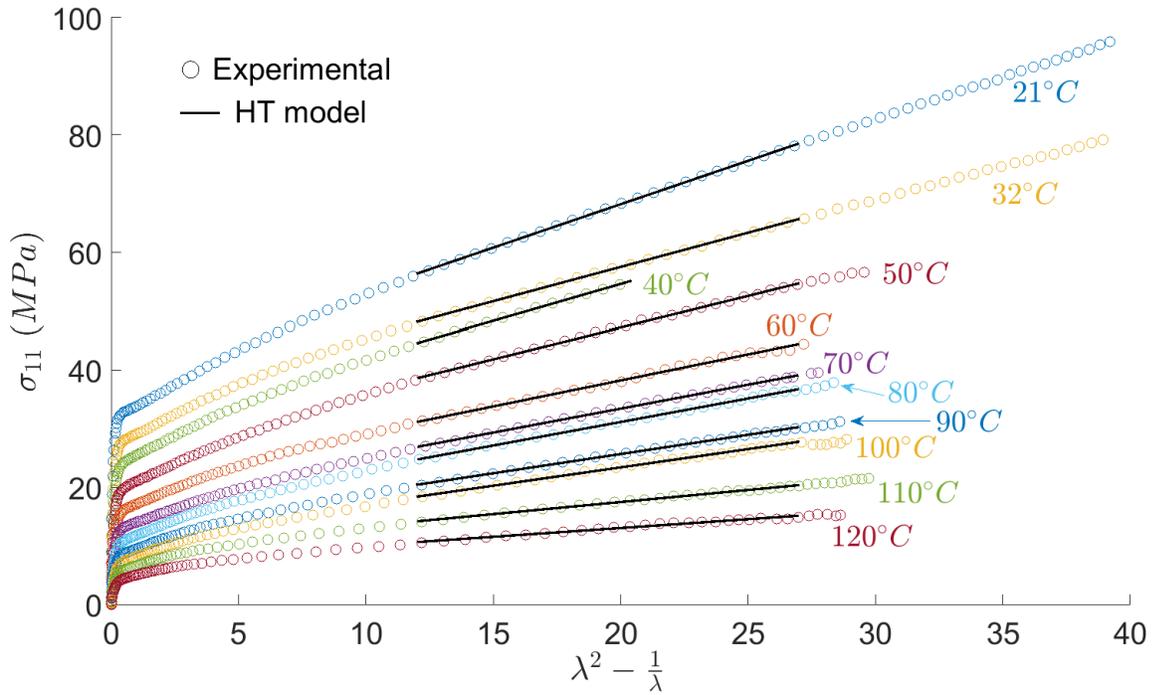

*Figure S6. $\sigma_{11} = Y + G(\lambda^2 - 1/\lambda)$ curves: experimental and HT model used to determine the $Y$ and $G$ parameters.*

Comments about figure S6 : For the sake of consistency, the HT model was applied to fit the data in the same $\lambda^2 - 1/\lambda \in [12-27]$ interval (except for $T = 40°C$ where no data are available because the image correlation algorithm did not work on the lateral face for $\lambda > 20$). However, it can be checked that if $\lambda > 27$, the linear HT model still works as can be seen on the curves obtained at $T = 21°C$ and $T = 32°C$ for which $\lambda^2 - 1/\lambda$ goes up to $40$.